# The ZK Spiral Family of Trajectories and Forces

J. West

Department of Chemistry and Physics, Indiana State University

January 8, 2024

**Abstract**

Particle trajectories in the form of a logarithmic spiral with specified angular time dependence, "ZK spirals," are shown to be analytic solutions for motion in non-central, but simple force power-laws. Each ZK spiral is a particular solution to a single associated force law. The position and velocity are determined as analytic functions of time in terms of the growth parameter, specified power law, and the initial conditions. Four examples, each for a different force law and relevant to the advanced classroom, are presented: the well-known attractive central inverse cube force; a bead on a rigid, horizontal, frictionless wire in the shape of the spiral trajectory; a car moving in a changing radius turn; and a known solution for a powered rocket with variable thrust in a Newtonian gravitational field. In this last case, general expressions for position and velocity of powered transfer orbits within the solar system are presented along with a new analytic expression for the rocket mass as a function of time or distance from the Sun. These results can be extended to include powered flight solutions to a generalized Lambert's problem for circular orbits in attractive central force laws of the form $F(r) = -F_o/r^q$, for q greater than or equal to 1.

**1. Introduction**

In physics, one usually knows the force(s) acting on an object, and the goal is to work towards a "solution" in the form of expressions for the position (and other parameters of interest) of the object as a function of time. There are important cases, however, when one already knows a trajectory, but would very much like to know the force that is responsible for that motion. The most famous problem of having a known trajectory, and looking for the force responsible is almost certainly the case of the motion of the planets about the Sun. The fact that the planetary orbits are elliptical allows one to show that the force of the Sun on the planets must obey a central inverse square law. That result can be extended to orbits that are conic sections (hyperbola, parabola and ellipse) which then also include repulsive Rutherford scattering.[1-4]

In addition, it is often not possible to find general solutions to a problem in analytical form, especially if the forces are non-linear. In such cases, obtaining even a single, "particular" solution in analytic form can be a significant step forward. A couple of examples relevant for the purposes of this paper include simple trajectories for specific **central** force laws: an offset circular orbit which passes through the origin (r = Rcosθ), is a particular solution to the force F = -$F_o/r^5$;[5-8] the cardioid orbit (r = R(1+ cosθ) is a particular solution to F = - $F_o/r^4$;[9] and an attractive central inverse cube force law with F = -$F_o/r^3$ allows a family of orbits, the Cotes' spirals.[10-12] The simplest and best known Cotes spiral is the logarithmic spiral (also called equiangular or growth spiral)[13] given in polar coordinates (r, θ) by r = $Re^{kθ}$, where R is the value of r at an angle of zero and k is the "growth parameter."[10-13]

Ion drive rocket propulsion systems are very low thrust, but very efficient, and have become widely used by aerospace concerns in the last couple of decades. Unlike the usual Hohmann orbital transfer maneuver where the engines are used impulsively and only at the start and end of the transfer, the ion thrusters are used over long stretches of the mission. The motion of a body only subject to an inverse square gravitational field already presents significant challenges to determine the time dependence of the motion. An additional force in the form of a thruster which provides a non-central and non-constant magnitude force would seem to make the problem intractable. It has in fact been known for some time that an analytic solution to motion in such a gravitational field with a variable thrust is possible, providing a complete time-dependent analytical solution in the form of the logarithmic spiral.[14]

In this paper, it is shown that logarithmic trajectories with a simple analytic expression for position as a function of time represent a particular solution to an entire "family" of force laws, with those trajectories and forces given by

$$\theta(t) \equiv B\ln(1 + Jt) \quad (1a)$$

$$r(\theta) \equiv Re^{k\theta} \quad (1b)$$

$$\vec{F}_{z,k}(r) \equiv (C\hat{r} + D\hat{\theta})r^z. \quad (1c)$$

The constant k is commonly known as the "growth parameter," z is any real number, and R is the distance from the origin at the angle zero. As will be shown below, the constants B, J, C, and D can all be determined in terms of z, k, and the initial speed $V_o$. For each choice of these parameters, one obtains only a single (particular) solution to a single, specific force law. Note that z = -3 is a notable exception to this rule, where the force is a central inverse cube force for all values of k. As z and k are such important parameters, the spirals, with the time dependence as specified in Eq. (1) will be referred to as "ZK spirals," to distinguish them from the more general term logarithmic spirals.

In general, z can be any real number, but four integer values of z will be chosen to provide physically relevant and potentially useful examples suitable for the advanced undergraduate classroom. The first is z = -3, which only allows D = 0, so that z = -3 represents the only central force law associated with the ZK spirals, a case that is well-known in central force motion studies, the central inverse cube force law. The second example is z = -1, which is mathematically equivalent to a bead sliding on a fixed, horizontal, rigid, frictionless wire bent into the shape of the trajectory. In this system, the torque applied to the object is constant, and the kinetic energy is constant. A potential energy function does exist but does not seem to provide much conceptual help.[15] Such constrained motion-on-frictionless wires or surfaces are found in a variety of instructional settings, such as an ice cube sliding off of an inverted sphere,[16] and frictionless motion along a cycloidal surface.[17] Constrained paths also serve pedagogical purposes for introducing students to computational methods. See especially reference 18 which includes expressions for velocity and acceleration for general wire shapes. The value of z = 0 is used for the third example, and it corresponds to a model of a car subject to simple friction, but driving on a track with a "changing radius" turn. The connection with simple friction, and recent investigations of constant magnitude dynamics makes this an attractive case for the classroom.

The last case to be considered is z = - 2. It is the most important, and the most extensive example, because it provides useful allowed analytic expressions for position and thrust as a function of time for powered trajectories in interplanetary space flight. For z = - 2, D is **not** zero, implying that a tangential force component is also required. Despite the combined influence of a gravitational field and a variable non-central thrust, the motion of the "variable thrust rocket" (VTR) is given in complete detail by Eq. (1). In modern rocket systems, the tangential component of the thrust is provided by ion thrusters that provide variable, low, but very efficient thrust. The thrust magnitude and direction are found from the expressions derived and that is critical information for VTR missions.[14,19,20] A new expression for the mass of the VRT, m(t) and/or m(r), is found in an analytic form and is similar to that of the famous "rocket equation."[21,22] The travel times obtained are compared to those found in previous investigations of Mars intercept missions using the more well-known Hohmann transfer orbits.[23-25]

While the ZK spiral is not an optimized "thrust profile" for the VTR, it is still very efficient **and** all aspects of the mission are known in terms of simple analytic expressions of time, making orbital dynamics calculations efficient enough to employ the much more efficient ion thrusters. The expressions obtained for z = -2 are also generalized to provide the equivalent expressions for powered flight in hypothetical central force laws of the form $F = F_o/r^q$ (q > -1), so that powered "transfer orbits" between the previously known closed orbits in inverse fourth power and inverse fifth power force fields could be examined using analytical functions of time for position and fuel usage.

Previous work investigating transfer orbits and Lambert's problem have provided algorithms that produce the time dependence of all of the orbits almost entirely in terms of analytic relationships. The portions of the calculations that require numerical methods can be performed in a spreadsheet.[26-29] Recent examples utilizing the Hohman transfer and Kepler orbits are suitable for the classroom.[22,30] However, using ZK spirals for the orbital transfer allows one to determine the travel times and fuel usage analytically, and the expressions for the ZK spirals can be generalized to powered rocket motion in attractive central forces of the form $F = -F_o/r^q$ with $F_o$ and q being constants with q > 1. This offers an opportunity to solve Lambert's problem for **any simple central attractive force law**, with analytic expressions for times and fuel costs.

The remainder of the paper is organized as follows: Section 2 presents the derivations of the expressions for position as a function of time, and the force laws as functions of z, k, and the initial conditions, Section 3 provides the example cases for z = -3, -1, 0, and -2, and conclusions are provided in Section 4.

## 2. ZK SPIRALS AND FORCE LAW FAMILIES

As a result of the exponential relationship in Eq. (1b) it is possible to determine many useful relationships of ZK spirals in relatively simple form. Consider the segment of one such spiral, as shown in Fig. 1. Notice that only a segment is shown, as the value of r is allowed to go to zero or to infinity in general. The path length S, traveled along the spiral is readily determined by integrating the path length and is found to be

$$S_{\theta=\theta_o,\theta_f} = \frac{R(1+k^2)^{1/2}}{k}\left|e^{k\theta_f} - e^{k\theta_o}\right|. \tag{2}$$

Trajectories that include the origin are of special interest as the particle will then pass through the force center. A particle traveling from r = 0 to r = R (or the reverse path) completes and infinite number of laps around the origin, but travels a finite total arc-length distance of[19]

$$S_{r=0,R} = \frac{R(1+k^2)^{1/2}}{k}. \tag{3}$$

This result is a general relationship of the trajectory of Eq. (1b), and does not depend on the specifics of the force law.

To determine the force causing a particular trajectory, it is necessary to know the first and second derivatives of the position (Eqs. (1a) and (1b)) with respect to time.

$$\theta(t) \equiv B\ln(1+Jt) \qquad\qquad r(\theta) = Re^{k\theta} = R(1+Jt)^{kB} \tag{4a}$$

$$\dot\theta = \frac{BJ}{1+Jt} \qquad\qquad \dot r = kBRJ(1+Jt)^{kB-1} \tag{4b}$$

$$\ddot\theta = \frac{-BJ^2}{(1+Jt)^2} \qquad\qquad \ddot r = RJ^2(kB)(kB-1)(1+Jt)^{kB-2}. \tag{4c}$$

These expressions, along with Eq. (3) show that a particle on an inward ZK spiral will reach the origin after completing an infinite number of laps around the origin, but having traveled the finite total distance S in a total (finite) time of T = 1/J.[19] This also means that such a particle on an outward motion trajectory will necessarily escape the force center, even for large negative values of z and r(0) arbitrarily small. The acceleration in polar coordinates is, by definition,

$$\vec a = (\ddot r - r\dot\theta^2)\hat r + (r\ddot\theta + 2\dot r\dot\theta)\hat\theta. \tag{5}$$

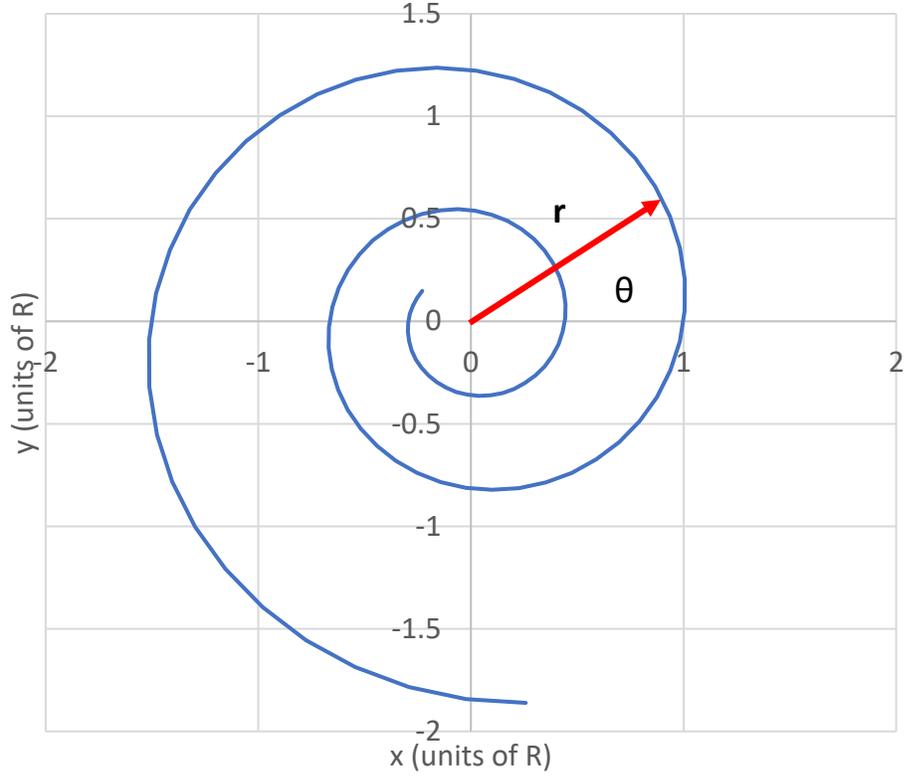

FIG. 1. A segment of the spiral for $r = Re^{k\theta}$, with $k = 0.130$, from theta = -10 radians to +5 radians. This segment has an arclength $S = 12.744\, R$, found via Eq. (2). The particle will always spiral in to the origin, or spiral outward to infinity, depending on which direction it is moving as the angle approaches negative or positive infinity.

Substituting from Eq. (4) into Eq. (5) gives the acceleration **a**, in terms of constants and the radial distance r

$$\vec{a} = RJ^2 \left(\frac{r}{R}\right)^{(kB-2)/kB} \{[kB(kB-1) - B^2]\hat{r} + [B(2kB-1)]\hat{\theta}\}. \quad (6)$$

It is convenient to eliminate B and J in favor of z, k, and the initial speed $V_o$, to give **F**

$$kB \equiv \frac{2}{1-z} \qquad\qquad J = \frac{kV_o(1-z)}{2R\sqrt{1+k^2}} \quad (7a)$$

$$V_o^2 = (1+k^2)(BJR)^2 \quad (7b)$$

$$\vec{F}_{z,k} = m\vec{a} = \frac{mV_o^2}{2R(1+k^2)}\left(\frac{r}{R}\right)^z \{[(1+z)k^2 - 2]\hat{r} + [k(3+z)]\hat{\theta}\}. \tag{7c}$$

The force is necessarily attractive for $z < -1$, but can be repulsive for $z > (2-k^2)/k^2$. The position, velocity components, kinetic energy KE, and angular momentum L, can all be written in terms of z, k, R and Vo,

$$\theta_{z,k}(t) = \frac{2}{k(1-z)}\ln\left[1 \pm \frac{V_o k|1-z|}{2R\sqrt{1+k^2}}t\right] \tag{8a}$$

$$r_{z,k}(t) = R\left[1 \pm \frac{V_o k|1-z|}{2R\sqrt{1+k^2}}t\right]^{2/(1-z)} \tag{8b}$$

$$V_\theta = \frac{V_o}{\sqrt{1+k^2}}\left(\frac{r}{R}\right)^{(1+z)/2} \qquad V_r = kV_\theta = \frac{kV_o}{\sqrt{1+k^2}}\left(\frac{r}{R}\right)^{(1+z)/2} \tag{8c}$$

$$KE = \frac{mV_o^2}{2}\left(\frac{r}{R}\right)^{(1+z)} \qquad |L| = \frac{mRV_o}{\sqrt{1+k^2}}\left(\frac{r}{R}\right)^{\frac{3+z}{2}} \equiv mrV_\theta. \tag{8d}$$

Here "+/-" indicates motion outward/inward relative to the origin for $z < 1$ (inward/outward for $z > 1$). If a particle moves along a ZK spiral defined by R, k, Vo, then the single value of z could be determined from observational time dependent data from Eq. (8a) or (8b). As mentioned before, Eq. (8b) shows that a particle on a ZK spiral will **always** reach the origin in a finite amount of time, or "escape" to an arbitrarily large value of r for large times.[19]

The angular momentum |L|, is conserved only for $z = -3$, which gives **the only purely central force law** associated with a ZK spiral, and the resulting force law is independent of k

$$\vec{F}_{-3,k} = m\vec{a} = -m\frac{V_o^2}{R}\left(\frac{r}{R}\right)^{-3}\hat{r}. \tag{9}$$

It is also possible to find ZK spirals associated with forces having only **purely tangential components** if $z > -1$ **and** k is restricted to a single value, k*, where

$$k^* = \sqrt{\frac{2}{1+z}} \qquad \vec{F}_{z,k*} = \frac{mV_o^2\sqrt{1+z}}{R\sqrt{2}}\left(\frac{r}{R}\right)^z\hat{\theta}. \tag{10}$$

Purely tangential forces of this form are known to produce logarithmic trajectories, but they are often expressed in terms of k rather than z.[31]

To finish this section, it should be noted that while $z = +1$ is an allowed solution, Eqs. (8a) and (8b) show that it is the "trivial solution" of the particle simply sitting, motionless, at the origin, so that $r(t) = 0$.

### 3. EXAMPLE CASES: z = -3, -1, 0, -2

The example cases will be restricted to forces with integer values of z. Such a restriction is not necessary, but these examples are relevant and mathematically suitable for the classroom.

### 3.1. Inverse Cube Central Force (z = -3)

From Eq. (7c), if $z = -3$, the force **must** be a central force. Using Eq. (1) and conservation of angular momentum gives a separable equation for the angular position

$$L_c = mr_c^2 \dot{\theta}_c \equiv mR^2\omega \tag{11a}$$

$$\dot{\theta}_c = \omega e^{-2k\theta_c}, \tag{11b}$$

where the subscript "c" indicates the central force case, and omega is the initial angular velocity. Integrating gives

$$\theta_c = \frac{1}{2k}\ln(1 + 2k\omega t) \qquad \dot{\theta}_c = \frac{\omega}{1 + 2k\omega t} = \omega e^{-2k\theta_c} \tag{12a}$$

$$r_c = R(1 + 2k\omega t)^{1/2} = Re^{k\theta_c} \qquad \dot{r}_c = \frac{kR\omega}{(1 + 2k\omega t)^{1/2}} = kR\omega e^{-k\theta_c}. \tag{12b}$$

Taking the time derivatives, then using Eq. (5), shows that the mass is moving in an attractive force field with $z = -3$, consistent with Eq. (9) for $V_o = \omega*R*\text{sqrt}(1+k^2)$. The results of Eq. (12) are consistent with the known previous solutions to this problem.[10]

### 3.2. A Bead on a Wire (z = -1)

Consider a bead sliding on a fixed, horizontal, rigid, frictionless wire bent into the desired spiral shape of Eq. (1b). The details of **r**(t) for the mass moving along the frictionless wire follow from Eq. (1b) and conservation of kinetic energy

$$E = \frac{mv^2}{2} = \frac{m}{2}\left(\dot{r}_w^2 + r_w^2 \dot{\theta}_w^2\right) = \frac{m}{2}\left(k^2 R^2 e^{2k\theta_w}\dot{\theta}_w^2 + R^2 e^{2k\theta_w}\dot{\theta}_w^2\right) \tag{13a}$$

$$\dot{\theta}_w e^{k\theta_w} = \left(\frac{2E}{mR^2(1+k^2)}\right)^{1/2} \equiv \omega \qquad\qquad \dot{\theta}_w = \omega e^{-k\theta_w}, \tag{13b}$$

where the subscript "w" has been used to indicate motion on the wire. The position is found by integrating with respect to time

$$\theta_w = \frac{1}{k}\ln(1+k\omega t) \qquad\qquad r_w = R e^{k\theta_w} = R(1+k\omega t). \tag{14}$$

The force on the bead is found from Eq. (6)

$$\vec{F}_{-1,k} = m\vec{a}_w = \frac{mR^2\omega^2}{r}(\hat{r}+k\hat{\theta}). \tag{15}$$

This is the only allowed force law family for which the kinetic energy is constant. The angular momentum $L_w$, of the bead and the torque on the bead are given by

$$L_w = mr_w^2 \dot{\theta}_w \qquad\qquad \tau_w = \frac{d}{dt}L_w = \frac{2Ek}{1+k^2}. \tag{16}$$

The net torque on the bead is constant for the duration of the motion. The curl of **F** is zero for z = -1, but the associated potential energy function does not seem to be very instructive.[15] The fact that the kinetic energy is conserved means that the path of the bead is along an equipotential line.

### 3.3. Variable Radius Racing Turns (z = 0)

This is the basic model of a car treated as a point mass and subject to static friction used in introductory courses (f = uN, where u is the coefficient of friction, and N is the "normal force"). The path, however, is **not** usually present in the introductory classroom, or on an "introductory" racetrack. A corner in the desired spiral shape is known in motor racing as a "changing radius turn." Such turns are notoriously challenging for drivers, in part, because they are of a different geometry than the overwhelming majority of those found on tracks.[32-34] The net force on the car is found from Eq. (7c)

$$\vec{F}_{0,k} = \frac{mV_o^2}{2R(1+k^2)}\{(k^2-2)\hat{r} + 3k\hat{\theta}\}, \tag{17}$$

which reduces, as expected, to uniform circular motion for k = 0. For z = 0, the direction of **F** is **not** constant (the second term depends implicitly on the angular position), but its magnitude **is** constant. The value of V$_o$ is fixed by the condition that the magnitude of the force is fixed, so that F$_{0,k}$ = umg. The value of k determines the shape of the turn and dictates the "racing line." The force **F**$_{0,k}$ has an attractive, repulsive, or no radial component, depending on the value of k. In racing applications, the shape of the track will usually dictate that k is less than 1, with a large attractive component.

The car continuously speeds up (slows down) on the increasing (decreasing) radius turn

$$V_\theta = \frac{V_o}{\sqrt{1+k^2}}\left(\frac{r}{R}\right)^{1/2} \qquad V_r = kV_\theta = \frac{kV_o}{\sqrt{1+k^2}}\left(\frac{r}{R}\right)^{1/2}. \tag{18}$$

The time to complete the turn is fixed by the values of k, R, V$_o$, and r. A spreadsheet can be used in the classroom to examine the effects of changing u on values of interest to the driver: exit speeds and the lap times. In a decreasing radius turn, **the car will reach a speed of zero and stop at the origin**. Likewise the car could start at the origin and participate in an outward spiral "drag race," similar to the motion of a car with constant magnitude force F = umg, on circular tracks. The motion on circular tracks is perhaps as challenging in the classroom as it requires some numerical work to obtain the values of interest.[35]

### 3.4. Variable Thrust Rocket

It would seem ambitious to attempt to solve for the position **r**(t) of a rocket that is moving in a Newtonian (1/r$^2$) gravitational field, while stipulating that the engines of the rocket provide a thrust that varies in magnitude and direction. However, it is well-known that the logarithmic spiral trajectory of Eq. (1b) is a solution for the motion of a rocket with thrust proportional to m/r², if the thrust is oriented at the correct constant angle relative to its position **r**, or velocity **v**, vector. Note, m is the "instantaneous mass of the spacecraft," m = m(t) or m = m(r).[14, 19]

Consider a mission to that most popular of hypothetical orbital destinations, Mars. At time t = 0 s the rocket is assumed to be at r(0) = R = 1.50 x 10¹¹ m = one AU from the Sun. Its initial tangential velocity is the same as the Earth in its orbit ($V_E$ = 2.974 x 10⁴ m/s). For a given value of k, the acceleration of the VTR can be found using Eq. (7c) with z = -2,

$$a_r = -\frac{GM}{r^2}\left(1 + \frac{k^2}{2}\right) \qquad a_\theta = +\frac{GM}{r^2}\left(\frac{k}{2}\right), \tag{19}$$

where G = 6.6743 x 10⁻¹¹ m³/(kgs²), the universal gravitational constant, and M = 1.989 x 10³⁰ kg, the mass of the Sun. The engines **do not** provide all force needed for the acceleration in Eq. (19). The engines are responsible **only for the terms containing "k."** A larger value of k requires higher values of acceleration and can dramatically increase fuel consumption, but will reduce the travel time. For small k, the ZK spiral is very similar to the trajectory obtained using a series of "patched conic orbits" produced by a series of impulsive burns, separated by long periods of ballistic free-flight paths (ref. 36, see Fig. 8).

For ease of comparison k is chosen so that the VRT reaches Mars after half an orbit around the Sun, the same as used in the traditional Hohmann transfer mission (see Fig. 2).[36,37] The desired value is k = 0.130, which gives $V_o$ = 1.0084$V_E$ = 2.999 x 10⁴ m/s. This value of is also similar to that used (k = 0.142) in the recent analysis of the motion of a solar sail by Bailer Jones (see Ref. 20, fig. 11, and Eq. (17)).

The VTR engines then need to provide a thrust $F_T$ with components found from Eq. (19) to be

$$F_{Tr} = -0.0085 \left(\frac{R}{r}\right)^2 \frac{GMm}{R^2} \qquad F_{T\theta} = 0.065 \left(\frac{R}{r}\right)^2 \frac{GMm}{R^2}. \tag{20}$$

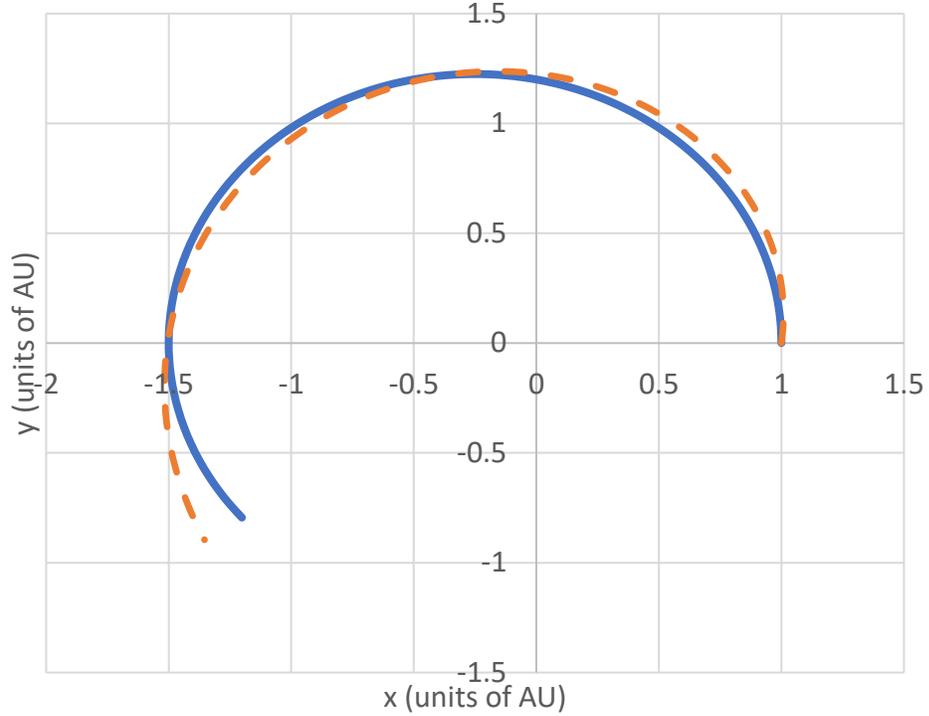

Figure 2. The trajectories of two orbital transfer missions around the Sun (at 0,0), between the Earth (r = R = 1 AU) and Mars (r = 1.5 AU). Both missions begin at r = R, and meet Mars on the opposite side of the Sun. The traditional elliptical Hohmann transfer is an ellipse (solid line). The ZK spiral for k = 0.130 (dashed line) has a radial velocity **away** from the Sun the entire mission. The travel times are found to be 0.709 yr for the Hohmann transfer and 0.693 yr for the ZK spiral.

The maximum thrust is required at r = R, where $|F_T| = 0.0655*(GMm/R^2) = m*(3.864 \times 10^{-4}$ m/s$^2$). The 3 ion engines of the Deep Space 1 mission provide a thrust that is on the order of 3 x 0.092 N = 0.276 N. That would be suitable for a VTR with a total (loaded) mass of m = 650 kg. This is well above the mass of the Deep Space One spacecraft (m = 373 kg "dry," and 486 kg fully loaded).[38,39] The ion engines are very efficient, and the electrical power and xenon fuel feed are indeed variable.[38-40] Very useful discussions of this class of propulsion systems can be found in reference 41, and the excellent review articles of references 42 and 43. The mission travel times are found in terms of r and R from Eq. (8b) with z = -2

$$T_{z=-2}(r,k) = \frac{\sqrt{1+k^2}}{k} \sqrt{\frac{R}{GM}} \left( \left(\frac{r}{R}\right)^{3k/2} - 1 \right). \qquad (21)$$

These are easily tabulated in a spreadsheet for quick mission parameter assessment in the classroom. The travel time for the Mars intercept at r = 1.5R is $T_{z=-2} = 2.186 \times 10^7 s = 0.693$ yr, approximately the same as that for the traditional Hohmann orbital transfer which takes 0.709 yr. This is to be expected given the very similar trajectories, as shown in figure 2 (compare with fig. 2 of ref. 23, and Eq. (36) of ref 19).

Another critical mission parameter is m(r), the mass remaining when the VTR arrives on station (the "payload"). Assuming, as usual, that the thrust provided by the engine is directly proportional to (exhaust velocity) * (rate fuel is expelled) and that the magnitude of the thrust varies as $m/r^2$ gives

$$ma_{thr} = -\frac{dm}{dt}V_{ex} \equiv \gamma \frac{m}{r^2}V_{ex}, \tag{22}$$

where $V_{ex}$ is the exhaust velocity, and the constant gamma is determined by matching the maximum thrust $F_T(r = R)$. Substituting r(t) into Eq. (22) gives a **separable** equation for m(t)

$$\frac{dm}{m} \equiv -\gamma \frac{dt}{r^2} = -\frac{\gamma}{R^2}(1+Jt)^{-\frac{4}{3}}.dt \tag{23}$$

On integrating, one obtains m(t). The constants J and gamma can be replaced in terms of the initial conditions, $V_{ex}$, and t can be replaced by r, giving

$$m_{VTR} = m_o exp\left\{-\left[\left(\frac{\sqrt{1+k^2}}{kV_{ex}}\right)(3.824 \times 10^3 \frac{m}{s})\right]\left[\frac{\sqrt{R}}{\sqrt{r_o}} - \frac{\sqrt{R}}{\sqrt{r}}\right]\right\}. \tag{24}$$

The parameter $m_o$ is the mass of the VTR when the main ion engines are first engaged. The mass remaining when arriving at Mars is found by setting r = 1.5R, and using $V_{ex}$ = 20,000 m/s

$$m_{VTR}(1.5R) = 0.760 m_o. \tag{25}$$

It should be noted that the estimate of m in Eq. (25) **is somewhat misleading**. It only accounts for the fuel used during the actual transit portion of the mission. The VTR must **first** switch from the circular orbit at r = R, with no radial velocity, onto the ZK spiral that has a radial velocity component of $k*V_E = 3.7 \times 10^4$ m/s (about 1/3 of escape velocity). An initial preparatory burn, using less efficient chemical rockets, is necessary to provide that initial radial velocity. Even then, the VTR uses much less fuel to reach Mars than a purely chemically powered equivalent mission.

The expression in Eq. (24) for m(r) can be generalized to VTR rockets operating in **central** force laws $F_c = -F_o/r^q$, for arbitrary q > 1 (z = -q) as

$$(m_{VTR})_q = m_o exp\left\{-\frac{2\gamma R^{1-n}\sqrt{1+k^2}}{kV_o(1-q)}\left[\left(\frac{r}{R}\right)^{(1-q)/2} - \left(\frac{r_o}{R}\right)^{(1-q)/2}\right]\right\}, \qquad (26)$$

where gamma is again determined in a similar manner as leading to Eqs. (19) and (20) by the maximum necessary acceleration. Transfer orbits for q = 3 (z = -3) are a special case that require **no fuel for the transfer portion of the mission**, as the ZK spirals **are** ballistic orbits in the inverse cube force law. Significant fuel is still needed to switch back and forth from circular orbits to ZK spirals, but the analytic form of the relationships makes the study of such missions very practical.

The attractive inverse fourth power, and inverse fifth power laws (q = 4 and q = 5) now offer especially interesting orbital transfer and Lambert's problem opportunities. Every attractive central force law allows simple circular orbits centered on the origin. Each of these force laws also allows a second family of well-known orbital shapes: the cardioid orbits r(θ) = R(1+cosθ), of q = 4, and the off-center circular orbits r(θ) = Rcos(θ), of q = 5.[5-9] A single **tangential** burn (very efficient) at each end of the transfer allows a VTR to switch back and forth between those specialty orbits and a ZK spiral transfer orbit. Such a transfer trajectory is shown for a pair of cardioid orbits in figure 3, and for a pair of off-center circular orbits in figure 4. In the examples shown in Figs. 3 and 4, the VTR would make the two tangential burns at the same angle relative to the fixed stars to aid in viewing, but any desired radius and/or final orbit orientation can be used in the transfer process.

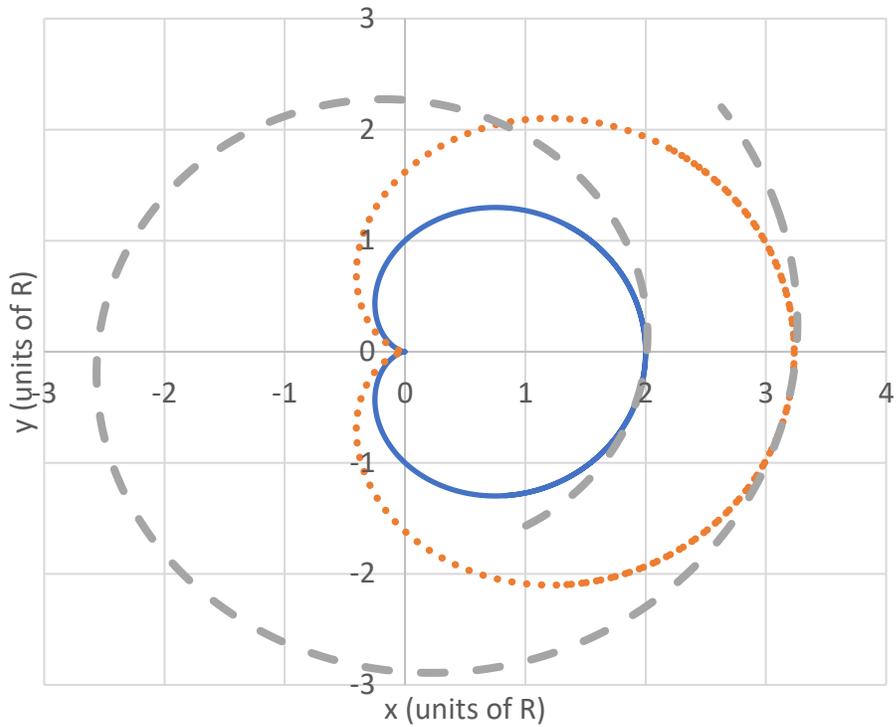

Figure 3. In an attractive inverse 4$^{th}$ power law, closed cardioid orbits of the form $r(\theta) = R(1+\cos\theta)$ are allowed. The ZK spiral transfer orbit (large dashes) with k = 0.07658 between the two cardioid orbits with R = 1 (solid line) and R = 1.618 (small dashes) is shown. For ease of visualization, k is chosen so that the ZK spiral intersects tangent to the cardioid orbits at the same absolute angle $\theta^*$, which satisfies the general relationship $(k^*)\sin(\theta^*) = 1+\cos(\theta^*)$. It is always possible to find a connecting ZK spiral, regardless of the relative orientations of the two cardioid orbits.

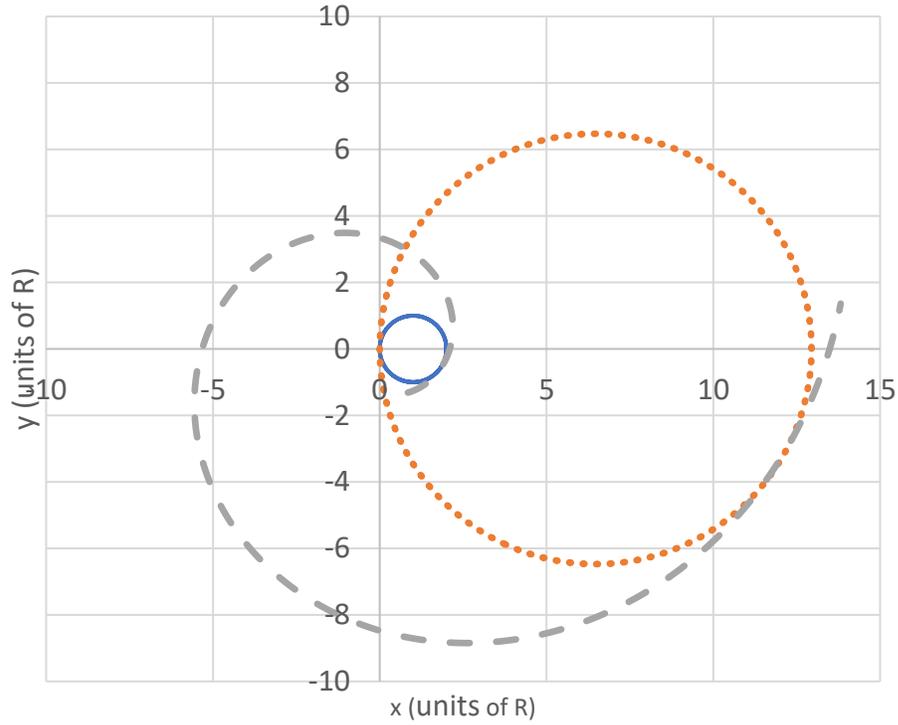

Figure 4. In an attractive inverse 5$^{th}$ power law, closed circular orbits of the form $r(\theta) = R\cos(\theta)$ are allowed. The ZK spiral transfer orbit (large dashes) with k = 0.296 connecting the off-set circular orbits with R = 1 (solid line) and R = 12.94 (small dashes) is shown. For ease of visualization, k is chosen so that the ZK spiral intersects tangent to the circular orbits at the same absolute angle $\theta^*$, which satisfies the general relationship $\tan(\theta^*) = -k^*$. It is always possible to find a ZK transfer orbit, regardless of the relative orientations of the two circular orbits.

**E. Trips to the Origin**

The examples conclude with calculations of the time required to reach the origin for the example cases already examined. For all cases, a particle on an inward moving ZK spiral will reach the origin after a time (see Eq. (8b))

$$t_{origin} = \frac{\sqrt{1+k^2}}{k(1-z)} \frac{2R}{V_o}. \tag{27}$$

Recall that the solution for z = +1 is simply r(t) = 0, and so Eq. (27) does not apply in that case and there is no danger of dividing by zero.

Consider a car, with tires allowing u = 1, on the ZK spiral: z = 0, k = 0.130, R = 100 m, $V_o$ = 31.3 m/s. With such a low value of k, the car is **almost** in uniform circular motion, so the reader might notice that $V_o$ is very close to sqrt (Rg)). The car will reach the origin in the time $t_{z=0}$ = 15.5 (R/$V_o$) = 49.6 seconds. A bead traveling on the **same** ZK spiral along on a rigid frictionless wire (z = -1) reaches the origin in half that time, $t_{z=-1}$ = ($t_{z=0}$)/2 = 24.8 s. Both objects begin at the same place and with the same initial speed, but the car slows down the entire trip, eventually stopping at the origin, while the bead on its frictionless wire maintains the same speed the entire time.

For z = -2, and z = -3, consider a pair of "Sun Diving" spacecraft on the same trajectory as the VTR mission to Mars (k = 0.130, R = 1.5 x $10^{11}$ m, $V_o$ = 3.024 x $10^4$ m/s) but on the reverse path. The VTR will reach the center of the Sun at $t_{z=-2}$ = 5.17 (R/$V_o$) = 0.813 yr, while the free-flying spacecraft in the central cubic force law will hit its equivalent of the Sun at the origin at $t_{z=-3}$ = (3/4)$t_{z=-2}$ = 3.88(R/Vo) = 0.610 yr.

## V. CONCLUSIONS

The new time dependent ZK spiral trajectories are allowed particular solutions for an infinite family of forces. Each spiral is a particular solution to a single associated force. Each force has radial and tangential components sharing a common dependence on the radial distance as $r^z$, where z is any real number. The force is necessarily attractive for z < -1, but can be attractive or repulsive for z >(2-$k^2$)/$k^2$. There are many useful applications for a force of this form in physics and the advanced undergraduate physics curriculum. Examples of ZK spirals corresponding to physical systems commonly encountered in upper-level undergraduate mechanics courses include z = -3 (the central inverse cube law), z = -1 (bead on a frictionless wire), z = 0 (driving on a changing radius turn), and z = -2 (powered space flight in an inverse square law) were presented.

The travel time and fuel used for a powered orbital transfer mission to Mars (z = -2) assuming an ion thruster drive was compared to the more traditional Hohmann transfer mission profile assuming chemical thrusters. The travel times are found to be almost equal, while the fuel usage of the ion thrusters is much lower. The ZK spiral mission profile can be extended to include powered rocket motion along ZK spirals in **any** central attractive force law of the form $F(r) = -F_q/r^q$ with $q > 1$. The mass of the rocket can be found in terms of a simple analytic function of time or radial distance for all values of q. This is in stark contrast to the mathematics normally associated with orbital dynamics, let alone the dynamics of a rocket using variable thrust engines in arbitrary central power law forces.

An investigation of Lambert's problem using ZK spirals in the usual Kepler application (q = 2) could prove interesting.[26-30] An advanced, optimization result has been formulated using "generalized logarithmic spirals" (three types, one of which is equivalent to the ZK spiral) for an Earth-Mars example is known.[44] One could also examine Lambert's problem for powered orbits in the q = 3, 4, and 5 force fields. It is easily shown that for q = 3 (attractive inverse cube force law) the spacecraft must always **reduce** its tangential speed when switching from circular orbits to a ZK spiral which would make the Lambert "chase" mission much more challenging than the Kepler (q = 2) equivalent. Analytic solutions for the intercept locations, times, and fuel used are possible, and it is expected that only minimal use of numerical methods should be required to solve only a single transcendental equation. The ZK spiral for z = +1 is the trivial solution r(t) = 0. It seems that some useful insight might also be gained by investigating the ZK spirals for z = 1 + d, where $|d| \ll 1$.

## AUTHOR DECLARATIONS
**Conflict of Interest**

The author has no conflict of interest to disclose.